%
%
%
\font\ninerm=cmr9
\font\eightrm=cmr8
\font\sixrm=cmr6
\font\ninei=cmmi9
\font\eighti=cmmi8
\font\sixi=cmmi6
\skewchar\ninei='177 \skewchar\eighti='177 \skewchar\sixi='177
\font\ninesy=cmsy9
\font\eightsy=cmsy8
\font\sixsy=cmsy6
\skewchar\ninesy='60 \skewchar\eightsy='60 \skewchar\sixsy='60

\font\ninebf=cmbx9
\font\eightbf=cmbx8
\font\sixbf=cmbx6
\font\ninett=cmtt9
\font\eighttt=cmtt8
\hyphenchar\tentt=-1 
\hyphenchar\ninett=-1
\hyphenchar\eighttt=-1
\font\ninesl=cmsl9
\font\eightsl=cmsl8
\font\nineit=cmti9
\font\eightit=cmti8
\newskip\ttglue
\def\tenpoint{\def\rm{\fam0\tenrm}%
  \textfont0=\tenrm \scriptfont0=\sevenrm \scriptscriptfont0=\fiverm
  \textfont1=\teni \scriptfont1=\seveni \scriptscriptfont1=\fivei
  \textfont2=\tensy \scriptfont2=\sevensy \scriptscriptfont2=\fivesy
  \textfont3=\tenex \scriptfont3=\tenex \scriptscriptfont3=\tenex
  \def\it{\fam\itfam\tenit}%
  \textfont\itfam=\tenit
  \def\sl{\fam\slfam\tensl}%
  \textfont\slfam=\tensl
  \def\bf{\fam\bffam\tenbf}%
  \textfont\bffam=\tenbf \scriptfont\bffam=\sevenbf
   \scriptscriptfont\bffam=\fivebf
  \def\tt{\fam\ttfam\tentt}%
  \textfont\ttfam=\tentt
  \tt \ttglue=.5em plus.25em minus.15em
  \normalbaselineskip=12pt
  \let\sc=\eightrm
  \let\big=\tenbig
  \setbox\strutbox=\hbox{\vrule height8.5pt depth3.5pt width0pt}%
  \normalbaselines\rm}
\def\ninepoint{\def\rm{\fam0\ninerm}%
  \textfont0=\ninerm \scriptfont0=\sixrm \scriptscriptfont0=\fiverm
  \textfont1=\ninei \scriptfont1=\sixi \scriptscriptfont1=\fivei
  \textfont2=\ninesy \scriptfont2=\sixsy \scriptscriptfont2=\fivesy
  \textfont3=\tenex \scriptfont3=\tenex \scriptscriptfont3=\tenex
  \def\it{\fam\itfam\nineit}%
  \textfont\itfam=\nineit
  \def\sl{\fam\slfam\ninesl}%
  \textfont\slfam=\ninesl
  \def\bf{\fam\bffam\ninebf}%
  \textfont\bffam=\ninebf \scriptfont\bffam=\sixbf
   \scriptscriptfont\bffam=\fivebf
  \def\tt{\fam\ttfam\ninett}%
  \textfont\ttfam=\ninett
  \tt \ttglue=.5em plus.25em minus.15em
  \normalbaselineskip=10pt 
  \let\sc=\sevenrm
  \let\big=\ninebig
  \setbox\strutbox=\hbox{\vrule height8pt depth3pt width0pt}%
  \normalbaselines\rm}
\def\eightpoint{\def\rm{\fam0\eightrm}%
  \textfont0=\eightrm \scriptfont0=\sixrm \scriptscriptfont0=\fiverm
  \textfont1=\eighti \scriptfont1=\sixi \scriptscriptfont1=\fivei
  \textfont2=\eightsy \scriptfont2=\sixsy \scriptscriptfont2=\fivesy
  \textfont3=\tenex \scriptfont3=\tenex \scriptscriptfont3=\tenex
  \def\it{\fam\itfam\eightit}%
  \textfont\itfam=\eightit
  \def\sl{\fam\slfam\eightsl}%
  \textfont\slfam=\eightsl
  \def\bf{\fam\bffam\eightbf}%
  \textfont\bffam=\eightbf \scriptfont\bffam=\sixbf
   \scriptscriptfont\bffam=\fivebf
  \def\tt{\fam\ttfam\eighttt}%
  \textfont\ttfam=\eighttt
  \tt \ttglue=.5em plus.25em minus.15em
  \normalbaselineskip=9pt
  \let\sc=\sixrm
  \let\big=\eightbig
  \setbox\strutbox=\hbox{\vrule height7pt depth2pt width0pt}%
  \normalbaselines\rm}
%
\def\headtype{\ninepoint}                 
\def\abstracttype{\ninepoint}             
\def\captiontype{\ninepoint}              
\def\footnotetype{\ninepoint}             
\font\chaptitle=cmr10 at 11pt             
\rm                                       

%
%
\parindent=0.25in                         
\parskip=0pt                              
\baselineskip=12pt                        
\hsize=4.25truein                         
\vsize=7.445truein                        
\hoffset=1in                              
\voffset=-0.5in                           

\newskip\sectionskipamount                
\newskip\aftermainskipamount              
\newskip\subsecskipamount                 
\newskip\firstpageskipamount              
\newskip\capskipamount                    
\newskip\ackskipamount                    
\sectionskipamount=0.2in plus 0.09in
\aftermainskipamount=6pt plus 6pt         
\subsecskipamount=0.1in plus 0.04in
\firstpageskipamount=3pc
\capskipamount=0.1in
\ackskipamount=0.15in
\def\sectionskip{\vskip\sectionskipamount}
\def\aftermainskip{\vskip\aftermainskipamount}
\def\subsecskip{\vskip\subsecskipamount} 
\def\firstpageskip{\vskip\firstpageskipamount}
\def\capskip{\hskip\capskipamount}

%
%
\nopagenumbers                            
\newcount\firstpageno                     
\firstpageno=\pageno                      
\newcount\chapno                          

\def\rightheadline{\headtype\phantom{\folio}\hfil\runningtitletext\hfil\folio}
\def\leftheadline{\headtype\folio\hfil\runningauthortext\hfil\phantom{\folio}}
\headline={\ifnum\pageno=\firstpageno\hfil
           \else
              \ifdim\ht\topins=\vsize           
                 \ifdim\dp\topins=1sp \hfil     
                 \else
                     \ifodd\pageno\rightheadline\else\leftheadline\fi
                 \fi
              \else
                 \ifodd\pageno\rightheadline\else\leftheadline\fi
              \fi
           \fi}

\def\bottomnumber{\hss\tenrm[\folio]\hss}
\footline={\ifnum\pageno=\firstpageno\bottomnumber\else\hfil\fi}

%
%
%
%
\outer\def\mainsection#1
    {\vskip 0pt plus\smallskipamount\sectionskip
     \message{#1}\vbox{\noindent{\bf#1}}\nobreak\aftermainskip\noindent}
 
\outer\def\subsection#1
    {\vskip 0pt plus\smallskipamount\subsecskip
     \message{#1}\vbox{\noindent{\bf#1}}\nobreak\smallskip\nobreak\noindent}
 
\def\backup{\nobreak\vskip-\baselineskip\nobreak\vskip-\subsecskipamount\nobreak
}

\def\title#1{{\chaptitle\leftline{#1}}}
\def\name#1{\leftline{#1}}
\def\affiliation#1{\leftline{\it #1}}
\def\abstract#1{{\abstracttype \noindent #1 \smallskip\vskip .1in}}
\def\ref{\noindent \parshape2 0truein 4.25truein 0.25truein 4truein}
\def\caption{\noindent \captiontype
             \parshape=2 0truein 4.25truein .125truein 4.125truein}

\def\footnote#1{\edef\fspafac{\spacefactor\the\spacefactor}#1\fspafac
      \insert\footins\bgroup\footnotetype
      \interlinepenalty100 \let\par=\endgraf
        \leftskip=0pt \rightskip=0pt
        \splittopskip=10pt plus 1pt minus 1pt \floatingpenalty=20000
        \textindent{#1}\bgroup\strut\aftergroup\strut\egroup\let\next}
\skip\footins=12pt plus 2pt minus 4pt 
\dimen\footins=30pc 

%
%

\def\@{\spacefactor 1000}

\def\,{\pcomma} 
\def\pcomma{\relax\ifmmode\mskip\thinmuskip\else\thinspace\fi}

\def\oversim#1#2{\lower0.5ex\vbox{\baselineskip=0pt\lineskip=0.2ex
     \ialign{$\mathsurround=0pt #1\hfil##\hfil$\crcr#2\crcr\sim\crcr}}}

\def\sigbar{ {\langle \sigma \rangle}}

\def\runningtitletext{The Stellar Initial Mass Function}
\def\runningauthortext{Meyer et al.}

\null
\firstpageskip

{\baselineskip=14pt
\title{The Stellar Initial Mass Function: Constraints from}
\title{Young Clusters and Theoretical Perspectives}
}

\vskip .3truein
\name{Michael R. Meyer}
\affiliation{Steward Observatory, The University of Arizona}
\vskip .2truein
\name{Fred C. Adams}
\affiliation{Department of Physics, The University of Michigan}
\vskip .2truein
\name{Lynne A. Hillenbrand, John M. Carpenter} 
\affiliation{Department of Astronomy, California Institute of Technology}
\vskip .2truein
\name{and Richard B. Larson}
\affiliation{Department of Astronomy, Yale University}
\vskip .3truein

\abstract{ 
We summarize recent observational and theoretical progress aimed at
understanding the origin of the stellar initial mass function (IMF)
with specific focus on galactic star--forming regions. 
We synthesize data from various efforts to determine the IMF 
in very young, partially--embedded stellar clusters and find: 
i) no significant variations in the low--mass IMF have been
observed between different star--forming regions; and 
ii) the mass distributions of young stars just emerging from molecular clouds
are consistent with having been drawn from the IMF derived from 
field stars in the solar neighborhood.  These results apply only to 
gross characterizations of the IMF (e.g. the ratio of high to low 
mass stars); present observations do not rule out more subtle regional 
differences.  Further studies are required in order to assess whether or 
not there is evidence for a universal turnover near the hydrogen--burning limit. 
We also provide a general framework for discussing 
theories of the IMF, and summarize recent work on several physical mechanisms
which could play a role in determining the form of the stellar 
initial mass function. 
}


\mainsection{{I}{}~~The Stellar Initial Mass Function}
\backup
A fundamental question in star formation is the origin 
of stellar masses.  Considerable progress has been made in recent years 
in understanding the formation of single stars, and we now have a
working paradigm of the process. Yet quantitative understanding of the 
distribution of stellar masses formed within molecular clouds remains elusive. 

One estimate of the initial distribution of stellar masses comes
from studies of volume--limited samples of stars in the solar neighborhood.  
Combining a variety of parallactic, spectroscopic, and
photometric techniques, a luminosity function is derived for main
sequence stars.  A main sequence mass--luminosity relationship is then
applied in order to derive the {\it present day mass function}, or
PDMF, from the luminosity function.  Next, the effects of stellar
evolution are taken into account in order to derive the {\it initial
mass function}, or IMF.  In constructing the ``solar neighborhood IMF,''  
the higher mass stars are typically drawn from young associations out to 
distances of a few kpc, whereas
the lower mass stars are drawn from well-mixed disk populations 
at distances out to tens of pc.  
Furthermore, because of their short main
sequence lifetimes, the high mass stars used in constructing the IMF
are quite young ($0.01-1 \times 10^8$ yr), while lower mass stars
found in the solar neighborhood are systematically older ($1-100
\times 10^8$ yr).  For these reasons, even a consistent definition of 
the IMF requires the assumption that it does not vary in time and space 
within the disk of the galaxy.  This implies 
that the molecular clouds which produced 
the stars currently found in the solar neighborhood each formed stars with the same IMF
as the regions producing stars today.  It is precisely this assumption 
we wish to test with the work reviewed below. 

The first comprehensive determination of the IMF over the full
range of stellar masses was given by Miller and Scalo (1979; hereafter
MS79), subsequently updated by Scalo (1986).  
Significant revisions have been made more recently at the low mass end
(e.g. Kroupa 1998; Reid 1998)
and at the high mass end (Garmany et al. 1982; Massey 1998).  
The reader is referred to Scalo (1986) for a detailed discussion of the 
ingredients that go into deriving the IMF and Scalo (1998) 
for a general review of the field.  Here we refer to the IMF 
as the number of stars per unit logarithmic mass interval, and will use
the power-law notation $\Gamma = d \log F(\log m_*)/d \log m_*$ to
characterize the IMF over a fixed mass range. 
In this notation, the slope of Salpeter (1955) is $\Gamma = -1.35$.

The IMF deduced from studies of OB associations and field stars in the 
solar neighborhood exhibits two main features that are generally agreed
upon.  First, for masses greater than about 5$M_{\odot}$, the IMF has
a nearly power-law form.  Massey (1995a; 1995b) find a slope $\Gamma =
-1.3 \pm 0.3$ for massive stars in clusters and a steeper power law
for high mass stars in the field.  Secondly, the mass function becomes
flatter for masses $< 1 M_{\odot}$ (Kroupa, Tout, \& Gilmore 1993; KTG).
Deriving the field star IMF near 1 M$_{\odot}$ is 
severely complicated by corrections for stellar evolution, which require
detailed knowledge of the star formation history of the galaxy.  Young
open clusters may be the best place to measure the IMF between 1--15
$M_{\odot}$ (e.g. Phelps \& Janes 1993).  At the lowest masses, there is considerable
debate whether or not the IMF continues to rise, is flat, or turns
over between $0.1-0.5M_{\odot}$ (Reid and Gizis 1997; Mera et
al. 1996; Kroupa 1995; Gould et al. 1997).  That the IMF
seems to change from a pure power--law to a more complex distribution
between 1--5 $M_{\odot}$ provides an important constraint on theories
for the origin of stellar masses.  Of even greater importance would be
the clear demonstration of a peak in the IMF at the low mass end.
Considerable observational effort has been focussed on establishing 
whether or not such a peak exists, and if so, characterizing 
its location and width (see Figure 1). 

In addition, it is extremely important to know whether or not the
time-- and space--averaged distribution of masses characterizing the
solar neighborhood is universal.  Do all star--forming events give rise
to the same distribution of stellar masses?  If star formation is
essentially a self-regulating process, then one might expect the IMF
to be strictly universal.  Alternatively, if stellar masses are
determined only by the physical structure of the interstellar medium
(e.g. fragmentation), then one might expect differences in
the IMF which depend on local conditions such as cloud temperature.
Certainly the details of either process depend on the physical
conditions of the clouds of gas and dust from which the stars form.
The unanswered question is: how sensitively does the distribution of
stellar masses depend on the initial conditions in the natal
environment?

In this contribution, we outline the observational and theoretical
progress which has been made toward understanding the origin of the
IMF in star--forming regions.  
In Section 2, we review advances in deriving mass distributions for 
very young clusters and attempt to summarize the ensemble of results. 
In Section 3, we review various theoretical constructs that have been
put forward to explain the expected shape of the IMF and its
dependance on initial conditions.  In Section 4, we summarize our
conclusions.

\mainsection{{I}{I}~~Stellar Mass Distributions in Star--Forming Regions}
\backup

Instead of providing a complete survey of the literature, we focus 
instead on a summary of approaches used and a comparison of results. 
We begin by describing why young clusters are 
helpful in understanding of the origin of the IMF. 
We then outline techniques employed to derive
stellar mass distributions in very young clusters, highlighting the
advantages and disadvantages of each method.  
Next, we describe several direct
comparisons of mass distributions assembled from the literature
on star--forming regions.
We then discuss a statistic which provides a gross characterization
of the IMF, the ratio of high-to-low-mass stars, and use data from
the literature to constrain its variation.
Finally, we compare work on star--forming regions with IMF studies 
of resolved stellar populations in the Milky Way and other galaxies. 

\subsection{A.~~The Utility of Young Clusters} 

Astronomers have long
used studies of galactic clusters to answer questions concerning the
formation and early evolution of stars (Clarke et al. and 
Elmegreen et al. this volume).  Bound open clusters provide 
a useful starting point, but because they are rare and long--lived 
it is thought that they do not contribute 
significantly to the field star population of the galactic disk
(cf. Roberts 1958).  In contrast, most young embedded clusters 
are thought to evolve into unbound associations, which comprise 
the majority of stars that populate the galactic disk (Lada \& Lada 1991).  
Embedded clusters are useful for studying the form of the IMF
    for several practical reasons.
First, because of their youth, evolutionary corrections needed to
translate the present day distribution of stellar masses into an
IMF are minimized.  Second, observations of such
clusters are more sensitive to low mass objects because the
mass--luminosity relationship for stars in the pre--main sequence
phase is not as steep a function of mass as for stars on the main
sequence.  Third, because of their compactness, they occupy small
projected areas on the sky, reducing the contamination by foreground
stars that plague studies of larger optically--visible associations.
Finally, the molecular cloud cores that contain embedded clusters
provide natural screens against background stars, which would
otherwise contaminate the sample.  Thus, we can attempt to derive
distributions of stellar masses for particular star--forming events
associated with individual molecular cloud cores. 
We can then determine whether or not the IMF varies 
as a function of cloud conditions, providing insight into its origin
(cf. Williams et al. this volume). 

\subsection{B.~~Approaches Used to Study Emergent Mass Distributions of Young Clusters}

Of course, partially--embedded young clusters also present difficulties to
astronomers interested in deriving their stellar mass distributions.
First of all, on--going star formation is typically observed in these
clusters.  As such, the observed mass distribution is only a
``snap--shot'' of the IMF for the star--forming event, and may not
represent the integrated final product of the cloud core.  In order to
keep this distinction clear between the {\it initial mass function} 
(as defined from the solar neighborhood sample) 
and the observed mass distributions of embedded young clusters, we will refer to these
``snap--shots'' of the IMF as {\it emergent mass distributions} (EMDs).
Another complication in studying embedded clusters is the time
dependent nature of the mass--luminosity (M--L) relationship 
(required for translating an observed luminosity function 
into a mass distribution) for PMS stars.  Third, large and
variable obscuration makes correction for extinction on an individual
star--by--star basis important for many embedded clusters.
Uncertainties in extinction--corrected absolute magnitudes can be
reduced by observing embedded clusters at longer wavelengths.  For
example, interstellar extinction in the K--band (2.2 $\mu$m) is $\times$ 10 
smaller than in the V--band (0.55 $\mu$m) and $\times$ 3 smaller
than in the J--band (1.25 $\mu$m).  However, at wavelengths beyond 2.0
$\mu$m, the observed flux from PMS cluster members is frequently
contaminated by excess emission.  The near--IR excess (associated with
hot circumstellar dust located in the inner--disk region) complicates
the interpretation of monochromatic near--IR luminosity functions as
{\it stellar} bolometric luminosity functions (e.g. Meyer et al. 1997).  
Most recent attempts to derive emergent mass distributions from
observations of embedded young clusters are based on: i) modeling of
monochromatic near--infrared luminosity functions; ii) analysis of
multi--color photometric data; iii) spectroscopic survey samples; or
iv) some combination of these techniques.  We discuss each of these 
methods (along with their pros and cons) below. 

A comprehensive review of embedded cluster work published before PPIII 
is given by Zinnecker, McCaughrean, and Wilking (1993; ZMW).  Many of these studies 
compared the observed distribution of K--band magnitudes with that expected 
from a standard IMF convolved with a main--sequence M--L 
relationship (e.g. Lada, Young, \& Greene 1992; Greene \& Young 1992).  
In contrast, ZMW constructed models for the
{\it evolution} of embedded cluster luminosity functions by transforming
theoretical PMS evolutionary tracks into an observational plane.  They 
generated synthetic K--band luminosity functions (KLFs) 
for clusters with ages 0.3--2 Myr from an assumed input IMF. 
Ali and DePoy (1995; see also Megeath 1996) made allowances for 
spatially--variable extinction and excess emission due to 
circumstellar disks in their analyses of KLFs. Lada and Lada 
(1995; see also Giovannetti et al. 1998) extended these approaches by 
considering continuous as opposed to discretized star formation.

With multi--color photometry, one increases the amount of information
and thus decreases the number of assumptions needed to derive an EMD.
Strom et al. (1993; see also Aspin et al. 1994) attempted to deredden 
individual embedded sources by
assuming the same intrinsic color for each star; they also considered
dereddened J--band luminosity functions (as opposed to K--band) to
minimize the effect of infrared excess emission.  Many studies of embedded 
young clusters assume an input mass distribution which is combined
with analysis of an observed luminosity function to derive the
cluster age or age distribution.  In principle, extensions of these
techniques could place constraints on the stellar mass distribution.
By making the shape of the input mass spectrum a variable, certain
distributions could be ruled out if found inconsistent with the
observed luminosity function for any reasonable input age distribution
(e.g. Lada, Lada, and Muench 1998).  In practice, treating both mass and
age distributions as variables makes it difficult to find a
unique best--fit solution.  

Comeron et al. (1993) developed a novel technique for analyzing multi--color 
photometric data of embedded clusters.  By modeling spectral energy distributions 
of embedded sources, one can estimate the exinction, intrinsic luminosities, 
and the spectral slopes of the SEDs.  The spectral slopes are 
taken as an indicator of evolutionary state and 
volume--corrections are applied to the sample as a function
of intrinsic luminosity.  This latter step takes into account the fact that 
more luminous objects probe larger volumes in 
flux--limited samples.  Adopting M--L relationships deemed appropriate
for the evolutionary state of the objects in question, the luminosity 
function is then transformed into an emergent mass distribution.  
Comeron et al. (1996) expanded on this approach by adopting appropriate M--L 
relationships in a Monte Carlo fashion from an assumed age distribution. 
Meyer (1996) developed an alternate method which utilizes multi--color photometry 
to deredden individual embedded sources, explicitly taking into account
the possibility of near--infrared excess emission.  Then by adopting an 
age (or age distribution) for a cluster, dereddened absolute J--band magnitudes are
used to estimate stellar masses.   In this case, an extinction--limited sample
is used to derive an emergent mass distribution that is not biased toward higher 
mass stars seen more deeply into the cloud. 

Ultimately, the construction of reliable IMFs for a large number of
young clusters, forming under a wide range of conditions, requires a
combination of deep photometric survey work and follow--up
spectroscopic analysis.  Spectra allow one to derive the photospheric
temperatures for the embedded objects and place them in the H--R
diagram.  Comparing the positions of sources in the H--R diagram with
PMS evolutionary models provides estimates of the mass and age
distributions of the sample.  Ideally, we would like to have {\it
complete} spectroscopic samples for each cluster. In practice, often the best 
we can do is obtain spectra for a sub--set of the photometric sample
(either representative or flux--limited), providing an estimate of the
cluster age.  With this constraint on the age--dependent M--L
relationship, one can characterize the dereddened luminosity function
in terms of the emergent mass distribution for the cluster and compare
it to the solar neighborhood IMF. 
While the spectroscopic techniques allow a more unambiguous
accounting for the effects of extinction, infrared excess and a stellar age
distribution, the photometric techniques have the distinct advantage that they
can be applied to fainter stellar populations at much greater distances.
It is important to remember that in all of these methods, the masses 
derived for individual stars depend
sensitively on the adopted PMS evolutionary tracks.  As a result,
estimates of mass scales at which inflections are observed in the
detailed distribution functions are necessarily uncertain until we
have a better calibration of PMS evolutionary tracks. 

\subsection{C.~~Three Techniques for Direct Comparison of EMDs}

Direct comparisons between observations of two
independent star--forming regions provide the best means to uncover
differences in the emergent mass distributions of clusters.
Statistical tests (such as the Kolmogorov--Smirnov test; Press et
al. 1993) between individual distributions are more informative than
comparisons to uncertain analytic functions such as those derived for
the field star IMF.  Provided that studies of young clusters are
performed in a uniform way, we can overcome uncertainties in the
various techniques by making relative rather than absolute
comparisons.  However, care must be taken in applying such statistical
tests, since any systematic differences in the observations between
two studies can easily produce a significant signal in the KS test.
Here, we restrict ourselves to three datasets, each of which has been
assembled using the same analysis so that meaningful comparisons for
different clusters can be made.
We begin with the KLF analyses of IC 348 and NGC 1333.
Next we review the multi--color photometric surveys of Ophiuchus and
NGC 2024, informed by follow--up spectroscopy.  Finally, we compare
the extensive spectroscopic surveys for the Orion Nebula cluster and
IC 348.

\medskip\noindent
{\it K--Band Luminosity Functions:  IC 348 vs. NGC 1333} 
\medskip 

IC 348 and NGC 1333 are both young clusters in the Perseus
cloud, located at a distance of 320 pc (Herbig 1998).
Lada and Lada (1995) present near--IR imaging of IC 348; 380 sources
are identified in excess of the measured field star population.
Because of low extinction and the low fraction of association members 
exhibiting strong near--infrared excess emission ($<$ 25 \%), 
the observed KLF gives a reasonable estimate of the stellar 
luminosity function, albeit convolved with an age distribution.  
Lada, Alves, and Lada (1996) 
performed a similar near--IR survey of NGC 1333, identifying a
``double--cluster'' of 94 sources.  However, in this case, differential
extinction seriously affects the observed KLF and, in addition,
$\sim$50\% of the association members displayed strong infrared excess emission.
To correct for extinction, the cluster populations were dereddened using
$(H-K)$ colors.  This dereddened KLF was compared with that observed for IC 348 
and with that published for the Trapezium cluster by ZMW, 
corrected for the difference in distance.
The KLF's for NGC 1333 and the Trapezium are consistent with having been drawn
from the same parent population.  As both clusters are thought to be the same 
age and to have similar IR excess frequencies, this result suggests that the
underlying mass functions are similar.  Comparison of the KLFs for NGC 1333 and
IC 348 yields a different result: there is only a 20 \% chance 
that they were drawn from the same distribution.  Lada, Alves, and Lada (1996) 
point out that the observed KLFs of all three clusters could be derived from the
same underlying mass functions, but convolved with different age
distributions.  Without independent estimates of the cluster
ages based on spectroscopic observations, it is difficult to draw
robust conclusions.

\medskip\noindent
{\it Combining Spectroscopy \& Photometry: Ophiuchus vs. NGC 2024} 
\medskip

Comeron et al. (1993; CRBR) conducted a K-band imaging
survey of the Ophiuchus cloud core (d $\sim$ 150 pc) 
within the $A_V > 50^m$ molecular contour and 
obtained follow--up four color photometry for all sources $K < 14.5^m$.  
Strom, Kepner, and Strom (1995; SKS) published a comparable multi--color 
imaging survey, focusing on the ``aggregates'' associated
with dense cores located within the $A_V > 50^m$ contour.  
Greene and Meyer (1995) combined infrared 
spectroscopy of 19 embedded sources with 
photometry from SKS in order to place them on the H--R diagram and to estimate 
masses and ages directly.  Williams et al. (1995) obtained spectra 
for three candidate low mass objects
from the sample of CRBR.  They compared the IMF results of CRBR and
SKS and found them to be in broad agreement.  Combining results from 
both studies of Ophiuchus yielded an emergent mass distribution from
0.1--5.0 $M_{\odot}$ with $\Gamma \sim -0.1$.

Comeron et al. (1996) performed a deep near--infrared survey of 
several fields toward the embedded 
cluster associated with NGC 2024 (d $\sim$ 470 pc). 
Assuming an age distribution for the cluster, they derive a mass 
function between 0.08--2.0 $M_{\odot}$ that is nearly flat with no
evidence for a turnover at the low mass end ($\Gamma = -0.2 \pm 0.1$)
consistent with the CRBR results for Ophiuchus. 
Meyer et al. (1999) present a multi--color near--IR survey of the
innermost 0.5 pc of NGC 2024, sampling 0.1 $M_\odot$ stars viewed
through $A_V < 19.0^m$.  Combining this photometric
survey with near--IR spectra for two dozen sources in the region, an
H--R diagram was constructed.  
Taking the age derived from the H--R diagram as characteristic of 
the entire embedded population, they
estimate the extinction toward each star and construct a 
stellar luminosity function complete down to 0.1 M$_{\odot}$.  

Given that existing near-infrared surveys of the embedded clusters
associated with NGC 2024 and the Ophiuchus molecular cloud are of
comparable sensitivity (M$_K < 4.5^m$ \& A$_V < 19^m$) 
and physical resolution($d \sim 400$ AU), we can directly
compare the derived luminosity functions.  
The K--S test reveals that there is a small chance that the stellar luminosity
distributions were drawn from the same parent population ($P = 0.04$).  
Meyer et al. (1999), in making this comparison, 
investigate three factors that can affect 
the shape of the stellar luminosity function for an embedded cluster: 
age distribution, accretion properties, and emergent mass distribution.
Comparison of the H--R diagrams shows
that the evolutionary states of both clusters are similar.  
Comparison of the JHK color--color diagrams for complete samples in
each cluster reveals that they are consistent with having been drawn
from the same parent population, suggesting that the accretion
properties of NGC2024 and Ophiuchus are also similar.  Although the KS test
suggests differences in the luminosity functions which could be
attributed to differences in the mass functions,
the uncertainties make it difficult to argue that the emergent mass 
distributions are significantly different. 

\medskip\noindent
{\it Complete Spectroscopic Samples:  IC 348 vs. The Orion Nebula Cluster}
\medskip

Herbig (1998) conducted an extensive photometric and
spectroscopic survey of IC 348, placing $\sim80$ optically--visible stars 
on the H--R diagram.  The derived age distribution was used to study
the mass distribution of a larger photometric sample in the $(V-I)$ vs V
color--magnitude diagram.  Based on a sample of 125 stars
fitted to the PMS evolutionary models of D'Antona and Mazzitelli (1994; DM94)
in the $(V-I)$ vs V
color--magnitude diagram, the derived mass function is similar to the
Scalo (1986) IMF down to 0.3 M$_{\odot}$.  
Luhman et al. (1998)
independently conducted an infrared and optical spectroscopic study of
the IC 348 region.  Spectra were obtained for 75 sources from the photometric
survey of Lada \& Lada (1995).  Luhman et al. claim
completeness of the spectroscopic sample down to 0.1 $M_{\odot}$.
Masses are estimated for the sources lacking
spectra by adopting the M--L relationship suggested by the extant
spectroscopic sample, and a corrected emergent mass distribution 
is constructed for the cluster.  
The Orion Nebula Cluster (ONC) is the richest young cluster 
within 1 kpc and has been the target of several photometric studies in the last
decade (Herbig and Terndrup 1986; McCaughrean \& Stauffer, 1994; Ali
\& DePoy 1995).  
Hillenbrand (1997) present results from an optical spectroscopic survey 
of nearly 1000 stars located within $\sim2$ pc of the Trapezium stars.  
From the resulting distribution of stellar masses 
derived from the H--R diagram (down to 0.1 $M_{\odot}$), Hillenbrand concludes
that the cluster mass function turns over at $\sim$ 0.2 M$_{\odot}$. 

IC 348 and the ONC have the most complete EMDs derived to date
among the very young clusters studied spectroscopically.  As a result,
it is desirable to compare them directly (Figure 2). 
We use masses derived from the DM94 tracks in this excercise, and note
that even this direct comparison using identical observational techniques
is only ``model independent'' to the extent that the clusters exhibit similar 
age distributions -- such that we are using comparable theory to translate
observables into stellar masses.  Comparison of the EMD 
derived globally for the ONC cluster (605 stars $A_V < 2.0^m$) with
that presented for IC 348 (73 stars $A_V < 5.0^m$) over the mass range
0.1--2.5 $M_{\odot}$ indicates that the two mass distributions were
not drawn from the same parent population ($P = 5.92 \times
10^{-5}$).  However, when a restricted sample from the inner ONC is
taken (133 stars within $r < 0.5$ pc, comparable to the
physical size of the Luhman et al. region in IC 348), we cannot rule
out that they were drawn from the same distribution ($P = 0.06$), 
despite $\times 100$ difference in central stellar density!  
This latter result could be due in part to the smaller sample considered; 
the smaller the differences between the EMDs of different regions, 
the larger the sample size needed to discern them.

\subsection{D.~~Synthesis of Results for an Ensemble of Clusters}

To study the IMF over the full range of stellar masses, we require 
photometric observations that are sensitive below the hydrogen burning limit 
for the distance and age of the cluster.  We also wish to sample the
stellar population through some well--determined and significant value
of extinction.  Finally, we need follow--up spectroscopy in order to 
inform cluster age estimates -- crucial for adopting an appropriate M--L 
relationship.   Why have so few quantitative results emerged from the 
study of EMDs in very young clusters?  Since PPIII, comprehensive 
spectroscopic studies of young stellar populations have been conducted 
towards a variety of star--forming regions (e.g. Alcala et al. 1997; 
Allen 1996; Hughes et al. 1994; Lawson et al. 1996; Walter et al. 
1994; 1997).  These studies, while providing crucial tests of many 
aspects of PMS evolution, are not ideally suited 
for studying emergent mass distributions. 
For example, H$\alpha$, x--ray, and variable
star samples impose activity--related selection effects 
which make it difficult to assess completeness.  Even in the well--studied 
Taurus dark cloud (Kenyon and Hartmann 1995) only now are samples 
complete down to the hydrogen burning limit becoming available (Briceno et al. 1998). 
Armed with deep photometric surveys, often we are still faced with a 
statistical problem.  The clusters and 
aggregates found in typical molecular clouds contain only tens to
hundreds of stars; populous regions like the ONC are rare.  As
demonstrated above, detailed comparisions of mass/luminosity functions
based on sample sizes $\ll 1000$ are inconclusive unless gross
differences exist in the underlying distributions.  In this section,
we compare results from a variety of recent studies and ask whether the
ensemble of results can tell us something about the shape of the IMF.

One approach is to use mass bins that are significantly wider than the
errors in the assigned stellar masses from the methods described above.
For example, in the study of nearby star-forming regions, a
particularly useful diagnostic is the ratio $\cal R$ of
intermediate-to-low-mass stars, 
$${\cal R} = N(1-10 M_{\odot}) / N(0.1-1M_{\odot}).$$ 
Each of the techniques outlined above for characterizing the emergent
mass distribution can be collapsed into such a ratio.  In Table I, we
present the $\cal R$ values for various regions for which the ratio of
intermediate-to-low-mass stars can be constructed, based on data
assembled from the literature.  We restrict this analysis to regions
located within 1kpc of the Sun, for which a flux--limited survey down
to a well--defined completeness limit exists, along with complementary
spectroscopy to constrain the age distribution of the association
members.  In Figure 3, we compare the ratio of intermediate-to-low-mass
stars derived for nine such regions to the same ratio predicted from
various analytic forms of the IMF. 

The range and errors in the values measured for the ratio ${\cal R}$
of intermediate-to-low-mass stars suggest a conservative conclusion:
{\it most extremely young, compact star--forming regions exhibit EMDs 
consistent with having been drawn from the field star IMF within our
ability to distinguish any differences.}  The small sizes of existing
observational samples, and the coarse nature of the tests we
have been able to perform, allow us to detect only gross
differences between the observed mass distributions and various forms 
of the field star IMF.  For example, it is clear that the Salpter
(1955) IMF does not hold below 1.0 $M_{\odot}$, as it predicts a ratio
of intermediate to low mass stars of ${\cal R}$ = 0.04, compared to
${\cal R}$ = 0.17 for the MS79 IMF.  Furthermore, the IMF does
not have a sharp truncation at masses well above 0.1 $M_{\odot}$.
If the MS79 IMF were truncated at 0.4 $M_{\odot}$, the
expected ratio of intermediate to low mass stars would be 0.4, which
is also excluded by the data.  Finally, based on the information in Table I, 
we can can rule out dramatic dependencies of the IMF on environmental 
characteristics such as cloud temperature (as measured from molecular 
line observations) or mean stellar volume density (averaged over a region 
$r \sim 0.3$ pc).

\subsection{E.~~Comparisons to Other Estimates of the Galactic IMF}

Because differences in observational technique could introduce
important systematic errors, it is dangerous (though interesting!) to
compare results derived for young clusters in star--forming regions
with other constraints on the IMF.  Much recent work has focussed on 
nearby open clusters with ages $<$ 1 Gyr which share some of the 
attributes that make star--forming regions excellent places to search 
for variations in the IMF.  Bouvier et al. (1998; see also
Williams et al. 1996; Zapatero Osorio et al. 1997) have derived the
IMF for the Pleiades open cluster well into the brown dwarf regime.
Although the ratios ${\cal R}$ for the Pleiades (120 Myr) and 
the much younger Trapezium are both consistent with having been drawn 
from the field star IMF, the detailed mass
distributions are much different ($P(d > obs) = 1.9 \times
10^{-38}$!).  This is probably due largely to differences in the
techniques used and the adopted PMS tracks, although it could reflect
true differences in the IMF.  
Hambly et al. (1995) studied the IMF below 1.0 M$_{\odot}$ 
in the 900 Myr--old metal--rich cluster Praesepe, finding a slope 
$\Gamma = -0.5$ between 0.1--1.0 M$_{\odot}$ though Williams et al. 
(1995a) find a somewhat shallower $\Gamma = -0.34 \pm 0.25$.  
Pinfield et al. (1997) 
have extended this work down into the brown dwarf regime and find 
a steeper rise in the mass function.  
There has been additional work identifying low mass 
stars and brown dwarf candidates in other clusters including 
$\alpha$ Per (Zapatero Osorio et al. 1996) and the twin clusters 
IC 2602/2391 (Stauffer et al. 1996), though detailed 
investigations of the IMF are not yet available. 

Phelps and Janes (1993) derive the IMF between 1.4--9.0 $M_{\odot}$
for a dozen older open clusters observed in the disk of the Milky Way.
Although most regions are consistent with a power--law slope of
$\Gamma = -1.4 \pm 0.13$, two regions exhibit significantly different
IMFs (NGC 581 \& NGC 663).  Similar results are reported by 
Sagar et al. (1996) though 
intriguing differences between clusters are found over small mass ranges.
Further support for a roughly universal
IMF comes from von Hippel et al. (1996), who demonstrate that the
observed turn--overs in luminosity functions observed in a variety of
cluster environments are correlated with metallicity (as expected,
given the dependence of the mass--luminosity relationship on
metallicity).  Studies of the low metallicity ``Pop
II'' component of the Milky Way are in surprisingly good agreement
concerning the shape of the low mass IMF.  Results from globular
clusters (cf. Chabrier \& Mera 1997; Cool et al. 1998 and references
therein), spheroid populations (Gould et al.
1997), and the bulge (Holtzman et al. 1998) indicate that $\Gamma
\sim 0.0 \pm 0.5$, consistent with the field star IMF and our results
for young clusters.  

Massey and collaborators (1998 and references therein) 
also find that the high mass end of the IMF does not
change with metallicity; their results suggest that the IMF is well
fit by a similar power--law in the LMC, the SMC, and the Milky Way.
These results are generally confirmed by other studies of resolved 
galactic and extragalactic OB associations
(e.g. Brown 1998; Bresolin et al. 1998). 
However, the mass distribution of ``field'' OB stars does seem to
differ from that for stars found in clusters.  In studies of the low
mass component of giant HII regions in the Milky Way (NGC 3603,
Eisenhauer et al. 1998; W3, Megeath et al. 1996) and the LMC (R136
in 30 Dor; Hunter et al. 1996; Brandl et al. 1996), the
distributions of stellar masses are consistent with the field star
IMF, at least down to 1.0 $M_{\odot}$.  Deeper observations obtained
at higher spatial resolution will be required to sample the IMF down
to the hydrogen--burning limit.  

Taken as a whole, the preponderance of the evidence suggests that crude 
characterizations of the IMF (such as the ratio of high to low mass stars)
do not vary strongly as a function of 
metallicity, star--forming environment, or cosmic time.  That said, there
are indications of possible variations in the IMF; the ONC vs. IC 348, 
results for NGC 581 \& 663 compared to other open clusters, and the massive 
star IMF derived from the field sample compared to results from associations. 
It is these differences which may provide 
clues to help unravel the mystery surrounding the origins of stellar masses. 
Finally, we note that all the results discussed
above are for the ``system IMF'', not accounting for unresolved binary
stars.  In order to quantify this effect, we need not only the
distribution of masses for the composite systems, but also the
distribution of companion mass ratios (Mathieu et al. this volume).  
Knowing both the IMF and the
distribution of companion mass ratios is crucial to understanding the
process of star formation.  However in what follows, we concentrate on
theories which might explain the origin of the system IMF.

\mainsection{{I}{I}{I}~~Some Theoretical Ideas on the Origin of the IMF}

Although the current theory of star formation remains incomplete, we
can begin to consider approaches to the IMF problem.  In this section,
we discuss briefly some current theoretical ideas that may be relevant
to the origin of the stellar IMF.  Given the limited space available,
we can only discuss some of the many ideas that have been proposed. 
We also emphasize that we still have only rather speculative ideas to
discuss.  Additional theoretical reviews can be found elsewhere 
(e.g. Clarke 1998, Larson 1998, Elmegreen 1998). 

\subsection{A.~~Towards a General Formulation}

Within the context of the current theory of star formation, it is
often useful to conceptually divide the process which determines the
IMF into two subprocesses: {\bf [1]} The spectrum of initial
conditions produced by molecular clouds or other star forming
environment.  {\bf [2]} The transformation between a given set of
initial conditions and the properties of the final (formed) star.

In this section, we outline one particular approach to constructing
theories of the initial mass function. In an IMF theory, the final
mass of a forming star must be determined through some physical
mechanism.  The identification of that mechanism, which could include
many different physical processes, lies at the heart of determining a
theory of the IMF. In practical terms, we need to specify the
transformation between the initial conditions and the final stellar
properties ( Adams \& Fatuzzo 1996; Khersonsky 1997). 
For example, the mass of the star $M_\ast$ could be given
by a ``semi-empirical mass formula'' (SEMF) or ``transfer function''
of the form $M_\ast = M_\ast (\alpha_1, \alpha_2, \alpha_3, \dots )$,
where the $\alpha_j$ are the physical variables that determine the
mass of the star. In the limit in which a large number of physical
variables is required to determine stellar masses, we obtain a
particularly simple result, suggesting a statistical
approach to the calculation of the IMF.

We would like to find a relationship between the distributions of the
initial variables and the resulting distribution of stellar masses
(the IMF).  For many (but certainly not all) cases of interest, 
the transformation can be written as a product of variables, i.e., 
$$M_\ast = \prod_{j=1}^N \alpha_j \, , \eqno(3.1)$$
where the $\alpha_j$ represent the relevant physical variables
(which could be the sound speed $a$, the rotation rate $\Omega$, etc., 
taken to the appropriate powers).  Each of these variables has a
distribution $f_j (\alpha_j)$ with a mean value $\ln {\bar \alpha}_j$ 
$\equiv$ $\langle \ln\alpha_j \rangle$ and a corresponding
variance $\sigma_j^2$.

In the limit of a large number $N$ of variables, the composite
distribution (the IMF) approaches a log-normal form.  This behavior is
a direct consequence of the central limit theorem (Richtmyer 1978) and
has been invoked by many authors (Larson 1973, Elmegreen \& Mathieu
1983, Zinnecker 1984, Adams \& Fatuzzo 1996).  Whenever a {\it large}
number of independent physical variables are involved in the star
formation process, the resulting IMF can be {\it approximately}
described by a log-normal form.  The departure of the IMF from a
purely log-normal form depends on the shapes of the individual
distributions $f_j$ and on the number of relevant variables.  However,
in the limit that the IMF can be described to leading order by a
log-normal form, there are simple relationships between the
distributions of the initial variables and the shape parameters $m_C$
and ${\langle \sigma \rangle}$ that determine the IMF. The mass scale
$m_C$ is determined by the mean values of the original variables
$\alpha_j$, and the dimensionless shape parameter $\sigbar$ is
determined by the widths $\sigma_j$ of the initial distributions, 
i.e., $m_C \equiv \prod_{j} \exp [ \langle \ln \alpha_j \rangle ]$ 
and  ${\langle \sigma \rangle}^2 = \sum_{j} \sigma_j^2$.  

In the limit that star formation involves a large number of
statistically independent variables, we would obtain a ``pure''
log-normal distribution.  In this limit, the only relevant parameters
are the total width of the distribution $\sigbar$ and the mass scale
of the distribution $m_C$, and these parameters are constrained by
observations of the IMF; The quantities $\sigbar$ and
$m_C$ are determined by the distributions of the physical variables in
the problem.  In a complete theory, we could calculate these initial
distributions from {\it a priori} considerations.  In the absence of a
complete theory, however, we can use observations of the physical
variables to estimate their distributions and hence determine
$\sigbar$ and $m_C$.

Although the number of physical variables involved in the star
formation process may be large, it is certainly not infinite.  An
important challenge of the future will be to unambiguously identify
the relevant physical variables in the problem.
In any case, the IMF will never 
{\it completely} converge to a log-normal form.  Instead, {\it the
distribution will retain tails}, departures from a log-normal form, 
at both the high mass and low mass ends.  Even though the composite
distribution does not obtain a purely log-normal form, however, the
theoretical predictions for the mass scale $m_C$ and total variance
$\sigbar$ must be consistent with the constraints from the observed
IMF.

\subsection{B.~~Gravitational Instability and the Jeans Criterion} 

In order for a clump of gas to collapse and form stars, its
self-gravity must overcome the effects tending to prevent collapse. 
On large scales, turbulence and magnetic fields provide the dominant
form of support against gravity, while in the densest core regions of
molecular clouds, thermal pressure is the dominant supporting force.  
A minimal requirement for collapse to occur is thus the classical Jeans
criterion that the self-gravity of a dense core must overcome its
thermal pressure; for a given temperature and density, this implies
a minimum size and mass called the Jeans length and Jeans mass.  For
density fluctuations in an infinite uniform medium with density
$\rho$ and isothermal sound speed $a$, the minimum unstable mass is
$\pi^{3/2}a^3\!/G^{3/2}\rho^{1/2}$ (Jeans 1929; Spitzer 1978).  Although
this result is not self-consistent in that it neglects the overall
collapse of the medium, dimensionally equivalent results are obtained
from analyses of the stability of equilibrium configurations.  For
a sheet, disk, or filament with surface density $\mu$, the minimum
unstable mass is a few times $a^4\!/G\mu^2$ (Larson 1985).  For an
isothermal sphere confined by an external pressure $P$, the minimum
unstable mass is $1.18\,a^4\!/G^{3/2}P^{1/2}$ (Ebert 1955; Bonnor 1956;
Spitzer 1968).  These expressions are all dimensionally equivalent
since $P = \rho a^2$ and since $P \sim G\mu^2$ in a self-gravitating
configuration; thus they are basically different expressions for the
same quantity, for which the term `Jeans mass' is a convenient name.  If
star-forming cores are created by turbulence in molecular clouds (Larson
1981), and if they are confined by a characteristic non-thermal pressure
arising from the cloud formation process (Larson 1996; Myers, 1998), 
then the best
estimate of the Jeans mass may be the mass of a critically stable
`Bonnor-Ebert sphere' bounded by this pressure.  For a typical molecular
cloud temperature of 10~K and a typical non-thermal cloud pressure of
$3 \times 10^5$ cm$^{-3}\,$K, this quantity is about  $0.7\,$M$_\odot$,
similar to the observed typical stellar mass (Larson 1998).

Direct evidence for the existence of gravitationally bound clumps
having about the Jeans size and mass in a star-forming cloud has been
found in recent millimeter continuum mapping of the $\rho$~Ophiuchus
cloud (Motte, Andr\'e, \& Neri 1998).  The observed clumps have masses
between 0.05 and $3\,$M$_\odot$ and a mass spectrum that closely
resembles the stellar IMF, becoming flatter below $0.5\,$M$_\odot$; thus
they may be the ``direct progenitors of individual stars or systems''
(Motte et al.\ 1998).  The separations between these clumps are
comparable to the predicted fragmentation scale of about $0.03\,$pc in
the $\rho$~Oph cloud, and this suggests that they have been formed by
gravitational fragmentation of the cloud.

Since a forming star grows in mass by accretion from a surrounding
envelope, the final mass that it attains depends on how much matter is
accreted.  Within the thermally supported inner region, the standard
picture of nearly radial infall from a nearly spherical envelope may
apply, but outside this region turbulence and magnetic fields dominate
the dynamics, and further accretion may be inhibited by these effects. 
A rigorous treatment of the problem is possible in the case where the
envelope is supported by a static magnetic field, as in the standard
model where collapse is initiated by slow ambipolar diffusion in a
magnetically supported configuration.  According to a number of recent
studies, even in this case a central region having approximately the
thermal Jeans mass begins to collapse dynamically at an early stage,
producing a high initial accretion rate but leaving behind a
magnetically supported envelope that is accreted more slowly (e.g.,
Basu 1997; Safier et al. 1997; Ciolek \& K\"onigl 1998).  Most of
the final stellar mass is probably acquired during the early phase of
rapid accretion, but infall may not stop completely after a Jeans mass
has been accreted.  The final termination of accretion may be caused by
the onset of a stellar wind, as discussed below. 

\subsection{C.~~The Role of Outflows}

In this section, we explore the possibility that stars themselves
determine their masses through the action of stellar winds and
outflows.  Protostellar outflows are sufficiently energetic to 
affect the overall support of a molecular cloud (Norman \& Silk 
1980) and have been conjectured to halt the inward accretion flow 
of a forming star (Shu 1985). This idea has been used as the basis 
for constructing various theoretical models of the IMF (e.g., 
Silk 1995, Nakano et al. 1995, Adams \& Fatuzzo 1996). 

Molecular clouds provide the initial conditions for the star forming
process.  These clouds are supported against gravity by both turbulent
motions and by magnetic fields.  As the fields gradually diffuse
outward, the clouds produce centrally condensed structures, molecular
cloud cores, which represent the initial conditions for protostellar
collapse.  In the simplest picture, these cores can be characterized
by two physical variables: the effective sound speed $a$ and the
rotation rate $\Omega$.  The effective sound speed generally contains
contributions from both magnetic fields and turbulence, as well as the
thermal contribution.

When molecular cloud cores undergo dynamic collapse, the central
regions fall in first and successive outer layers follow as pressure
support is lost from below (Shu 1977).  Because the initial state
contains angular momentum, some of the infalling material collects
into a circumstellar disk surrounding the forming star. The collapse
flow is characterized by a mass infall rate ${\dot M} \approx a^3/G$,
the rate at which the central star/disk system gains mass from the
collapsing core. The total amount of mass available to a forming star
is generally much larger than the final mass of the star.

In this rotating accretion flow, the ram pressure of the infall is
weakest at the rotational poles of the object.  The central star/disk
system gains mass until it is able to generate a powerful stellar wind
which breaks through the infall at the rotational poles and thereby
leads to a bipolar outflow configuration.  Although the mechanism
which generates these winds remains under study (e.g.  Shu et al.; 
K\"onigl \& Pudritz this volume), the characteristics of outflow sources have
been well determined observationally (Padman et al. 1997; 
Richer et al. this volume). 
The basic working hypotheses of the ``outflow conjecture'' is that
these outflows help separate nearly formed stars from the infalling
envelope and thereby determine, in part, the final masses of the
stars.  In this scenario, the transformation between
initial conditions and stellar masses is accomplished through the
action of stellar winds and outflows.  The central star/disk system
gains mass at a well--defined mass infall rate.  As the nascent star
gains mass over time, it becomes more luminous, and produces an increasingly
more powerful stellar outflow.  
When the strength of this outflow becomes larger than the
ram pressure of the infalling material, the star separates itself from
the surrounding molecular environment and thereby determines its final
mass.

We can use this idea to calculate a transformation between the initial
conditions in a molecular cloud core and the final mass of the star
produced by its collapse.  Using the idea that the stellar mass is
determined when the outflow strength exceeds the infall strength, we
can write this transformation (the SEMF) in the form 
$$L_\ast M_\ast^{2} = 8 m_0 \gamma^3 \delta
{\beta \over \alpha \epsilon}  {a^{11} \over G^3\Omega^2}
= \Lambda \, {a^{11} \over G^3\Omega^2}  \, , \eqno(3.2)$$
where the parameters $\alpha$, $\beta$, $\gamma$, $\delta$, and
$\epsilon$ are efficiency factors (Adams \& Fatuzzo 1996; Shu et
al. 1987).  This formula specifies a transformation between
initial conditions (the sound speed $a$ and the rotation rate
$\Omega$) and the final properties of the star (the luminosity
$L_\ast$ and the mass $M_\ast$).  Since the protostellar luminosity
$L_\ast$ as a function of mass is known, we can find the final stellar
mass in terms of the initial conditions.  In general, all of the
quantities on the right hand side of equation [3.2] will have a
distribution of values.  These individual distributions ultimately
determine the composite distribution of stellar masses $M_\ast$.

In principle, the physical variables appearing in equation [3.2] can
be measured observationally.  If we use the observed distributions of
these variables to estimate the shape parameters appearing in the IMF,
we obtain $m_C \approx 0.25$ and $\sigbar \approx 1.8$. These values
are in reasonably close agreement with those required to fit the MS79
IMF, namely $m_C$ = 0.1 and $\sigbar$ = 1.57.  Although a quantitative
comparison with observations is premature, this approach to the IMF
contains some predictive power and is roughly consistent with
observations.

The main weakness of this outflow approach is that the interaction
between stellar outflows and the inward accretion flow has not yet
been calculated. Highly collimated outflows can only reverse the
infall along the poles of the system. The outflows must therefore widen
with time and must be able to suppress accretion over most of the
solid angle centered on the star. Although some observational evidence
suggests that outflows can successfully reverse the inward accretion
flow (Velusamy \& Langer 1998), this issue remains open, on both the
theoretical and observational fronts.

\subsection{D.~~Hierarchical Fragmentation}

Since the Jeans mass decreases with increasing density, it is possible
that a collapsing cloud can fragment into successively smaller pieces
as its density increases.  This hypothesis of hierarchical
fragmentation (Hoyle 1953) has formed the basis of many theories of
the IMF.  In principle, a wide spectrum of stellar masses can be
produced in this way, depending on how many clumps stop subdividing
and collapse directly into individual stars at each stage of the
overall collapse.  For example, if the probability of further
subdivision of a cloud fragment is the same for each unit logarithmic
increase in density, a log-normal IMF is produced (Larson 1973).

Numerical simulations of fragmentation in collapsing clouds suggest
that hierarchical fragmentation is of limited importance.  For
example, a rotating cloud does not fragment significantly until it has
collapsed to a disk, since the development of subcondensations is
inhibited by pressure gradients and does not get ahead of the overall
collapse (Tohline 1980; Monaghan \& Lattanzio 1991).  This result
reflects the inconsistency of the original Jeans analysis, which
neglected the overall collapse, and suggests that significant
fragmentation does not occur until large-scale collapse has stopped
and a near-equilibrium configuration has formed.  Such a configuration
may then fragment as expected from linear stability theory (Larson
1985).  The formation of an equilibrium disk is an idealized case, but
transient near-equilibrium structures such as sheets or filaments may
often be created by turbulence in molecular clouds.  Much of the
observed structure of these clouds is in fact filamentary, and many
observed star-forming cores may have formed by the fragmentation of
filaments (Schneider \& Elmegreen 1979).  Thus, star-forming cores may
form directly by a single stage of fragmentation, rather than through
a series of stages of hierarchical fragmentation.

Hierarchical fragmentation may still be relevant to the formation of
low-mass stars in the binary and small multiple systems that are the
typical outcome of the collapse of Jeans-mass cloud cores (Larson
1995).  Numerical simulations show that multiple systems containing
several accreting and interacting `protostars' are often formed in
this case (Burkert, Bate, \& Bodenheimer 1997), and these objects are
often surrounded by disks or spiral filaments that can fragment into
yet smaller objects, so that some amount of hierarchical fragmentation
is possible.  However, coalescence can also occur and reduce the
number of fragments, and more work is needed to determine the final
outcome (Bodenheimer et al. this volume).  
Such small-scale fragmentation processes may be responsible
for the formation of the least massive stars, and may help to
determine the form of the lower IMF.  At present, however, both
observations and simulations suggest that only a small fraction of the
mass participating in such processes goes into the smallest objects or
`proto-brown-dwarfs'.

We can place this hierarchical fragmentation scenario into the general
picture of \S 3.A. As a cloud core fragments into successively smaller
pieces, we can follow one particular chain of fragmentation events to
the final fragment mass.  After one round of fragmentation, the piece
that we are following has mass $M_1 = f_1 M_0$, where $f_1$ is a
fraction of the original core mass $M_0$.  After $N$ iterations of
this hierarchy, the fragment mass will be $M_N$.  Although one could
identify the final fragment mass $M_N$ with the mass of the star
$M_\ast$ formed therein, in practice, the final fragment is not yet a
star. Instead, it provides the initial conditions for the star
formation process.  To account for the additional physical processes
that occur as the fragment collapses into a star, we can write 
$M_\ast = f_E M_N = M_0 f_E f_1 f_2 \dots f_N$, where $f_E$ is the
star formation efficiency factor of the final fragment.  We thus
obtain another SEMF and \S 3.A applies. For example, the total
variance of the distribution is given by $\sigbar^2 = \sigma_E^2 +
\sum \sigma_j^2$ where the $\sigma_j$ are the variances of the
distributions of the fragmentation fractions $f_j$.

\subsection{E.~~Accretion and Agglomeration Processes}

Star formation could also involve continuing accretion or
agglomeration processes, especially for the most massive stars which
typically form in dense clusters containing many less massive stars. 
Jeans fragmentation seems unlikely to be relevant to the formation
of these stars, and some kind of accumulation process may instead be
required (Larson 1982).  Radial accretion of gas is inhibited for very
massive stars because of the effects of radiation pressure (Wolfire et
al.\ 1985), but rotation may allow infalling material to collect into
a disk which can then accrete onto the star (Jijina \& Adams 1996). 
Alternatively, the material that builds massive stars may already be
in the form of dense clumps which may then accumulate into larger
objects.  An extreme case of this scenario would be the merging of
already formed stars (Stahler, this volume).

No convincing prediction of the form of the upper IMF has yet
emerged from any of these ideas, but since accretion or agglomeration
processes have no preferred scale, they could in principle proceed in
a scale-free manner and build up the observed power-law upper IMF.  As a simple
example, if each star accretes matter from a diffuse medium at a rate
proportional to the square of its mass, the upper IMF becomes a power
law with a slope not very different from the original Salpeter form
(Zinnecker 1982).  Models involving the agglomeration of randomly moving
clumps have also been studied, and these models can yield approximate
power-law mass functions (e.g., Nakano 1966; Silk \& Takahashi 1979;
Pumphrey \& Scalo 1983; Murray \& Lin 1996), but their predictions are sensitive to the
assumed properties of the clumps and their evolution between collisions.
Fractal concepts have been invoked to relate the form of the upper IMF
to the possible self-similar structure of molecular clouds (Larson 1992;
Elmegreen 1997), but in these theories strong assumptions must be made
about how matter accumulates into stars at each level of the fractal
hierarchy.  Finally, it has been suggested that already formed stars
might sometimes merge to form more massive stars, a mechanism that would
readily overcome the problems posed by radiation pressure and winds
(Bonnell, Bate, \& Zinnecker 1998; Stahler, this volume).  The dynamics
of systems dense enough for frequent mergers to occur is likely to be
chaotic, but since no new mass scale is introduced, such processes might
proceed in a scale-free way and build up a power-law upper IMF.

In summary, accretion and agglomeration processes almost certainly
play an important role in the formation of the most massive stars,
and they could plausibly proceed in a scale-free way and build up a
power-law upper IMF.  However, a quantitative understanding of these
processes does not yet exist, and much more work will be needed before
a reliable prediction of the slope of the upper IMF is possible.

\bigskip 

\mainsection{{I}{V}~~Summary of Conclusions and Future Directions}

In this contribution, we have reviewed the techniques 
used to analyze recent observations of mass distributions 
in very young clusters and attempted to summarize results.  
We have also endeavored to review 
current theoretical arguments that have been put forth to explain
these observations.  While the observational results are still not
definitive, several clear trends are emerging: 

\noindent 
[1]  Detailed comparisons of emergent mass distributions for the best 
studied young clusters suggest that the IMF does not vary wildly from 
region to region though more subtle differences could still be uncovered. 

\noindent 
[2] The ensemble of results which characterize the mass distributions
of embedded clusters, such as the ratio of intermediate to low mass
stars, is consistent with having been drawn from the field star IMF 
and {\it rule out} a single power--law 
Salpeter IMF that extends from 0.1--10 $M_{\odot}$. 

\noindent
[3] Although the evidence remains preliminary, the emergent mass
distributions derived for the two best studied young clusters (IC 348 
and the ONC) hint at turnovers between 0.1 -- 0.5 $M_{\odot}$.

Currently, several limitations prevent us from drawing more robust 
conclusions. Pre-main sequence evolutionary tracks remain uncertain
and hence we cannot determine stellar masses with the requisite
accuracy.  We also need to measure the distribution of companion mass
ratios to properly account for unresolved binaries, especially for the
low mass end of the IMF.  Next, we must search for variations in the 
substellar mass function, as well as extreme environments (starburst
analogs etc.) in order to test whether the IMF is truly universal.
Improved statistical methodology is also needed to help push our data to the natural
limits imposed by information theory.  Finally, it will soon become
possible to combine detailed IMF studies in young clusters with 
determinations from other resolved stellar populations in the Milky Way 
and other galaxies in order to search for variations over cosmic time.

Theoretical progress is being
made in understanding the origin of the IMF, but the problem is formidable. 
Stellar masses are determined by a complex interplay between the
initial conditions in the natal cloud environment and the stars that
are formed therein.  In this review, we have suggested a general
framework for discussing theories of the IMF, and we have outlined 
briefly some of the physical mechanisms that come into play including
gravitational instability and the Jeans scale, the effect of outflows,
hierarchical fragmentation, and accretion or agglomeration processes.
While it is still too premature to make 
meaningful comparisons between theories and 
observations of the IMF, it is our 
expectation that in the coming years
such a comparison can be effected. 

\vskip 0.5in

{\it We thank C. Lada and G. Rieke for
reading an earlier version of this work, 
and an anonymous referee for a critical review
of the manuscript. 
MRM, LAH, and JMC would like to express their gratitude to 
S. Strom for encouraging them along the road, paved 
with good intentions, that led to our 
interest in the IMF.}

\vfill\eject
\null 

\vskip .5in
\centerline{\bf REFERENCES}
\vskip .25in

\ref{Adams, F. C., \& Fatuzzo, M. 1996.  A Theory of the Initial Mass Function 
for Star Formation in Molecular Clouds.  ApJ, 464, 256. }

\ref{Alcala, J.M., Krautter, J., Covino, E., Neuhauser, R., Schmitt, J, \& Wichmann, R. 
1997.  A Study of the Chamaeleon Star--Forming Region from the ROSAT All--Sky Survey.  
II.  The Pre--Main Sequence Population.  A\&A, 319, 184.}

\ref{Ali, B., \& Depoy, D. 1995.  A 2.2 Micron Imaging Survey of the 
Orion A Molecular Cloud.  AJ , 109, 709.}

\ref{Allen, L.E. 1996.  Star Formation in Lynds 1641.  PhD Thesis, University of Massachusetts.}

\ref{Aspin, C., Sandell, G., \& Russell, A.P.G. 1994.  Near--IR Imaging Photometry of NGC 1333:
I.  The Embedded PMS Stellar Population.  A\&AS, 106, 165.}

\ref{Basu, S. 1997.  A Semianalytic Model for Supercriticial Core Collapse: 
Self--similar Evolution and the Approach to Protostar Formation.  ApJ, 485, 240.}

\ref{Bonnor, W.B., 1956.  Boyle's Law and Gravitational Instability.  MNRAS, 116, 351.}

\ref{Brandl, B., Sams, B.J., Bertoldi, F., Eckart, A., Genzel, R., Drapatz, S., 
Hofmann, R., Loewe, M., and Quirrenbach, A. 1996.  Adaptive Optics Near--Infrared 
Imaging of R136 in 30 Doradus:  The Stellar Population of a Nearby Starbust.  ApJ, 466, 254.} 

\ref{Bresolin, F., Kennicutt, R.C., Ferrarese, L., Gibson, B.K., Graham, J.A., Macri, L.M., 
Phelps, R.L., Rawson, D.M., Sakai, S., Silbermann, N.A., Stetson, P.B., Turner, A.M. 1998.
A Hubble Space Telescope Study of Extragalactic OB Associations.  AJ, 116, 119.}

\ref{Brown, A. 1998.  The Initial Mass Function in Nearby OB Associations. 
The Stellar Initial Mass Function, eds. G. Gilmore \& D. Howell (San Francisco:  ASP)} 

\ref{Bodenheimer, P., Ruzmajkina, T.,  \& Mathieu, R.D. 1993.  Stellar Multiple Systems: 
Constraints on the Mechanism of Origin.  
Protostars and Planets III, eds. E. Levy \& J. I. Lunine 
(Tucson: Univ. Arizona Press).}

\ref{Bonnell, I.A., Bate, M.R, \& Zinnecker, H. 1998.  On the Formation of Massive Stars. 
MNRAS, 298, 93.}

\ref{Bouvier, J., Stauffer, J.R., Martin, E.L, Barrado, 
Y. Navascues, D., Wallace, B., \& Bejar, V. J. S. 1998. 
Brown Dwarfs and Very Low--Mass Stars in the Pleiades Cluster: 
A Deep Wide--field Imaging Survey. A\&A, 336, 490.}

\ref{Brice\~no, C., Hartmann, L., Stauffer, J.  \& Martin, E. 1998. 
A Search for Very Low Mass Pre--Main Sequence Stars in Taurus. 
AJ, 115, 2074.}

\ref{Brown, A.G.A., de Geus, E.J., \& de Zeeuw, P.T. 1994. 
The Orion OB1 Association:  I.  Stellar Content. A\&A, 289, 101.}

\ref{Burkert, A., Bate, M. R., \& Bodenheimer, P., 1997. 
Protostellar Fragmentation in a Power--Law Density Distribution. MNRAS, 289, 497.}

\ref{Burkert, A. \& Bodenheimer, P. 1996.
Fragmentation in a Centrally--Condensed Protostar.  MNRAS, 280, 1190.}

\ref{Carpenter, J.M., Meyer, M.R., Dougados, C., Strom, S.E., 
\& Hillenbrand, L.A. 1997.  
Properties of the Monoceros R2 Stellar Cluster.  AJ, 114, 198.}

\ref{Chabrier, G., Mera, D. 1997.  Determination of the Globular Cluster
and Halo Stellar Mass Functions and Stellar and Brown Dwarf Densities.  
A\&A, 328, 83.}

\ref{Clarke, C. 1998.  Star Formation Theories and the IMF. The Stellar Initial Mass Function, 
eds.  G. Gilmore \& D. Howell (San Francisco: ASP).} 

\ref{Ciolek, G.E. \& K\"onigl, A. 1998.  Dynamical Collapse of Nonrotating Magnetic 
Molecular Cloud Cores:  Evolution Through Point--Mass Formation.  ApJ, 504, 257.}

\ref{Comeron, F., Rieke, G.H., Burrows, A., \& Rieke, M.J., 
1993.  The Stellar Population in the $\rho$ Ophiuchi Cluster. 
ApJ, 416, 185.}

\ref{Comeron, F., Rieke, G.H., \& Rieke, M.J. 1996.
Properties of Low--Mass Objects in NGC 2024.  ApJ, 473, 294.}

\ref{Comeron, F., Rieke, G.H., Claes, P., Torra, J., Laureijs,
R.J. 1998.  ISO Observations of Candidate Young Brown Dwarfs. 
A\&A, 335, 522.}

\ref{Cool, A.M. 1998.  Measuring Globular Cluster Mass Functions with HST.
The Stellar Initial Mass Function, eds.  
G. Gilmore \& D. Howell (San Francisco: ASP).}        

\ref{D'Antona, F. \& Mazzitelli, I. 1994.  New Pre--Main Sequence Tracks for 
$M_* < 2.5 M_{\odot}$ as Tests of Opacities and Convection Model. 
ApJS, 90, 467.}

\ref{Ebert, R., 1955.  Temperatur des interstellaren Gases bei 
grossen \break 
Dichten.  Z. Astroph., 37, 217.}

\ref{Eisenhauer, F., Quirrenbach, A., Zinnecker, H., \& Genzel, R.  1998.
Stellar Content of the Galactic Starburst Template NGC 3603 from Adaptive 
Optics Observations.  ApJ, 498, 278.}

\ref{Elmegreen, B.G. 1998.  Observations and Theory of the Stellar 
Initial Mass Function.  Unsolved Problems in Stellar Evolution, ed. M. Livio
(Cambridge:  Cambridge University Press).} 

\ref{Elmegreen, B.G. 1997.  The Initial Stellar Mass Function from 
Random Sampling in a Turbulent Fractal Cloud. ApJ, 486, 944.}

\ref{Elmegreen, B. G., \& Mathieu, R. D. 1983. 
Monte Carlo Simulations of the Initial Stellar Mass Function.  MNRAS, 203, 305.}

\ref{Garmany, C.D., Conti, P.S., \& Chiosi, C. 1982.
The Initial Mass Function for Massive Stars.  ApJ, 263, 777.}

\ref{Giovannetti, P., Caux, E., Nadeau, D., \& Monin, J. 1998. 
Deep Optical and Near--Infared Imaging Photometry of the Serpens Cloud Core. 
A\&A, 330, 990.} 

\ref{Gould, A., Flynn, C., \& Bahcall, J.N. 1998.
Spheroid Luminosity and Mass Functions from Hubble Space Telescope Star Counts. 
ApJ, 503, 798.}

\ref{Gould, A., Bahcall, J.N., \& Flynn, C. 1997. 
M Dwarfs from Hubble Space Telescope Star Counts:  III.  The Groth Strip 
ApJ, 482, 913.}

\ref{Greene, T.P. \& Young, E.T. 1992.
Near--Infrared Observations of Young Stellar Objects in the 
$\rho$ Ophiuchi Dark Cloud.  ApJ, 395, 516.}

\ref{Greene, T.P. \& Meyer, M.R. 1995.
An Infrared Spectroscopic Survey of the $\rho$ Ophiuchi 
Young Stellar Cluster:  Masses and Ages from the H--R Diagram. 
ApJ, 450, 233.}

\ref{Hambly, N.C., Steele, I.A., Hawkins, M.R.S., \& Jameson, R.F. 1995.
The Very Low--Mass Main Sequence in the Galactic Cluster Praesepe. 
MNRAS, 273, 505.} 

\ref{Herbig, G. 1998.  The Young Cluster IC 348.  ApJ, 497, 736.}

\ref{Herbig, G.H. \& Terndrup, D.M. 1986. 
The Trapezium Cluster of the Orion Nebula.  ApJ, 307, 609.}

\ref{Hillenbrand, L.A.  1997.  On the Stellar Population and Star--Forming 
History of the Orion Nebula Cluster.  AJ, 113, 1733.}

\ref{Hillenbrand, L.A., Meyer, M.R., Strom, S.E, \& Skrutskie, M.F.
1995.  Isolated Star--Forming Regions Containing Herbig Ae/Be Stars:
I.  The Young Stellar Aggregate Associated with BD+404124.  AJ, 109, 280.}

\ref{Holtzman, J.A., Watson, A.M., Baum, W.A., Grillmair, C.J., Groth,
E.J., Light, R.M., Lynds, R., \& O'Neil Jr, E.J., 1998.
The Luminosity Function and Initial Mass Function in the Galactic Bulge. 
AJ, 115, 1946.}

\ref{Hoyle, F., 1953.  On the Fragmentation of Gas Clouds into Galaxies and Stars. 
ApJ, 118, 513.} 

\ref{Hughes, J., Hartigan, P., Krautter, J., \& Keleman, J. 1994.
The Stellar Population of the Lupus Clouds.  AJ, 108, 1071.}

\ref{Hunter, D.A., Vacca, W.D., Massey, P., Lynds, R., \& O'Neil,
E.J. 1996.  Ultraviolet Photometry of Stars in the Compact Cluster
R136 in the Large Magellanic Cloud. 
AJ, 113, 1691.}

\ref{Jeans, J.H., 1929, Astronomy \& Cosmogony (Cambridge:  Cambridge 
University Press).} 

\ref{Jijina, J. \& Adams, F.C., 1996.
Infall Collapse Solutions in the Inner Limit:  Radiation Pressure and Its Effects
on Star Formation.  ApJ, 462, 874.} 

\ref{Jones, T.J., Mergen, J., Odewahn, S., Gehrz, R.D., Gatley, I., 
Merrill, K.M., Probst, R., \& Woodward, C.E. 1994.  A Near--Infrared 
Survey of the OMC2 Region.  AJ, 107, 2120.} 

\ref{Kenyon, S.J., \& Hartmann, L. 1995.  Pre--Main Sequence Evolution in 
the Taurus--Aurgia Molecular Cloud.  ApJS, 101, 117.} 

\ref{Khersonsky, V. K. 1997.  The Connection Between the Interstellar Cloud 
Mass Spectrum and the Stellar Mass Spectrum in Star--Forming Regions.  ApJ, 475, 594.}

\ref{Kroupa, P., Tout, C.A., \& Gilmore, G. 1993.  The Distribution of 
Low--Mass Stars in the Galactic Disc.  MNRAS, 262, 545.}

\ref{Kroupa, P. 1995.  Unification of the Nearby and Photometric 
Stellar Luminosity Functions.  ApJ, 453, 358.} 

\ref{Kroupa, P. 1998.  The Stellar Mass Function.  Brown Dwarfs and Extra--solar Planets, eds. 
R. Rebolo, E. Martin, \& M. Zapatero--Osorio (San Francisco: ASP).} 

\ref{Lada, C.J., Young, E.T., \& Greene, T.P. 1993.  Infrared Images of the Young 
Cluster NGC 2264.  408, 471.}  

\ref{Lada, C.J., Alves J., Lada E. A., 1996.  Near--Infrared Imaging of 
Embedded Clusters:  NGC 1333.  AJ, 111, 1964.} 

\ref{Lada, C.J., \& Lada, E.A, 1991.  The Nature, Origin, and Evolution 
of Embedded Star Clusters.  The Formation and Evolution of Star Clusters, 
ed. K. Janes (San Francisco:  ASP).} 

\ref{Lada, E.A., \& Lada, C.J, 1995.  Near--Infrared Images of IC 348 and the 
Luminosity Functions of Young Embedded Star Clusters.  AJ, 109, 1682. } 

\ref{Lada, E., Lada, C.J., \& Muench, A., 1998.  Infrared Luminosity Functions of 
Embedded Clusters.  The Initial Mass Function, 
eds. G. Gilmore \& D. Howell (San Francisco: ASP).}

\ref{Larson, R. B., 1973.  A Simple Probabilistic Theory 
of Fragmentation.  MNRAS, 161, 133.} 

\ref{Larson, R. B., 1981.  Turbulence and Star Formation in Molecular Clouds.  MNRAS, 194, 809.} 

\ref{Larson, R. B., 1982.  Mass Spectra of Young Stars. MNRAS, 200, 159.} 

\ref{Larson, R. B., 1985.  Cloud Fragmentation and Stellar Masses. MNRAS, 214, 379.} 

\ref{Larson, R. B. 1991.  Some Processes Influencing the Stellar Initial 
Mass Function.  Fragmentation of Molecular Clouds and 
Star Formation, ed. R. Capuzzo-Dolcetta (Dordrecht: Kluwer).} 

\ref{Larson, R. B., 1992.  Towards Understanding the Stellar 
Initial Mass Function.  MNRAS, 256, 641.} 

\ref{Larson, R. B., 1995.  Star Formation in Groups.  MNRAS, 272, 213.} 

\ref{Larson, R. B., 1996.  Star Formation and Galactic Evolution. 
The Interplay Between Massive Star Formation, the ISM, and Galaxy Evolution,  
eds. D. Kunth, B. Guiderdoni, M. Heydari-Malayeri, and T. X. Thuan
(Gif Sur Yvette: Editions
Frontieres), 3.}

\ref{Larson, R. B., 1999.  Theoretical Aspects of Star Formation. 
The Orion Complex Revisted, eds. M.J. McCaughrean \& 
A. Burkert (San Fransico:  ASP).} 

\ref{Lawson, W.A., Feigelson, E.D., \& Huenemoerder, D.P. 1996.  An 
Improved H--R Diagram for Chamaeleon I Pre--Main Sequence Stars. 
MNRAS, 280, 1071.} 

\ref{Luhman, K. \& Rieke, G.H., 1998.  The Low--Mass Initial Mass 
Function in Young Clusters:  L1495E.  ApJ, 497, 354.} 

\ref{Luhman, K., Rieke, G.H., Lada, C.J., \& Lada, E.A. 1998.
Low--Mass Star Formation and the Initial Mass Function in IC 348.  ApJ, 507, 347.} 

\ref{Massey, P., Lang, C., DeGioia-Eastwood, K. \& Garmany,
C.D. 1995a.  Massive Stars in the Field and Associations of the 
Magellanic \break 
Clouds:  The Upper Mass Limit, the Initial Mass Function, 
and a Critical Test of Main--Sequence Stellar Evolutionary Theory. 
ApJ, 438, 188.}

\ref{Massey, P., Johnson, K.E., \& DeGioia-Eastwood, K., 1995b.  The Initial 
Mass Function and Massive Star Evolution in the OB Associations of the 
Northern Milky Way.  ApJ, 454, 151.}

\ref{Massey, P. 1998.  The Initial Mass Function of Massive Stars in the 
Local Group.  The Stellar Initial Mass Function, eds. G. Gilmore \& D. Howell 
(San Francisco: ASP).} 

\ref{McCaughrean, M.J. \& Stauffer, J.R., 1994.  High Resolution Near--Infrared 
Imaging of the Trapezium:  A Stellar Census.  AJ, 108, 1382} 

\ref{McKee, C. F., Zweibel, E. G., Goodman, A. A., \& Heiles, C. 1993.
Magnetic Fields in Star--Forming Regions:  Theory.
Protostars and Planets III, eds. E. Levy \& J. I. Lunine 
(Tucson: Univ. Arizona Press), 327.}

\ref{Megeath, S.T. 1996.  The Effect of Extinction on the K--Band
Luminosity Functions of Embedded Stellar Clusters.  A\&A, 311, 135.} 

\ref{Megeath, S.T., Herter, T., Beichman, C., Gautier, N., 
Hester, J.J., Rayner, J., \& Shupe, D. 1996.  A Dense Stellar 
Cluster Surrounding W3 IRS 5.  A\&A, 307, 775.} 

\ref{Mera, D., Chabrier, G., \& Baraffe, I. 1996.  Determination of the Low--Mass 
Star Mass Function in the Galactic Disk.  ApJ, 459, L87.} 

\ref{Meyer, M.R., 1996.  Stellar Populations of Deeply Embedded Young Clusters: 
Near--Infrared Spectroscopy and Emergent Mass Distributions. 
PhD Thesis, University of Massachusetts.} 

\ref{Meyer, M.R., Calvet, N., \& Hillenbrand, L.A. 1997. 
Intrinsic Near--Infrared Excesses of T Tauri Stars:  Understanding the 
Classical T Tauri Locus.  AJ, 114, 288. } 

\ref{Meyer, M.R., Carpenter, J.M., Strom, S.E., \& Hillenbrand,
L.A. 1999.  The Embedded Cluster Associated with NGC 2024.  AJ, submitted.}

\ref{Miller, G.E., \& Scalo, J.M.  1979.  The Initial Mass Function and 
Stellar Birthrate in the Solar Neighborhood.  ApJS, 41, 513} 

\ref{Motte, F., Andr\'e, P., \& Neri, R., 1998.  The Initial Conditions
of Star Formation in the $\rho$ Ophiuchi Main Cloud:  Wide--Field 
Millimeter Continuum Mapping.  A\&A, 336, 150.} 

\ref{Monaghan, J.J. \& Lattanzio, J.C. 1991.  A Simulation of the Collapse and 
Fragmentation of Cooling Molecular Clouds.  ApJ, 375, 177.}

\ref{Mouschovias, T., 1991.  Magnetic Braking, Ambipolar Diffusion, Cloud 
Cores, and Star Formation -- Natural Length Scales and Protostellar Masses. 
ApJ, 373, 169.}

\ref{Murry, S. D., \& Lin D.N.C. 1996.  Coalescence, Star Formation, and the 
Cluster Initial Mass Function.  ApJ, 467, 728.}

\ref{Myers, P. C. 1998.  Cluster--Forming Molecular Cloud Cores.  ApJ, 496, L109.}

\ref{Nakano, T., Hasegawa, T., \& Norman, C. 1995.  The Mass of a Star Formed in 
a Cloud Core:  Theory and Its Application to the Orion A Cloud.  ApJ, 450, 183.}

\ref{Nakano, T. 1966.  Fragmentation of a Cloud and the Mass 
Function of Stars in Galactic Clusters.  Prog. Theor. Physics, 36, 515.}

\ref{Norman, C. A., \& Silk, J. 1980.  Clumpy Molecular Clouds - 
A Dynamic Model Self-Consistently Regulated by T Tauri Star Formation. 
ApJ, 238, 158.}  

\ref{Phelps, R.L. \& Janes, K., 1993.  Young Open Clusters as Probes 
of the Star--Formation Process:  2.  Mass and Luminosity Functions of 
Young Open Clusters.  AJ, 106, 1870.} 

\ref{Pinfield, D.J., Hodgkin, S.T., Jameson, R.F., Cossburn, M.R., 
\& von Hippel, T. 1997.  Brown Dwarf Candidates in Praesepe. MNRAS, 287, 180.} 

\ref{Press, W., Teukolsky, S.A., Vetterling, W.T., \& Flannery, B.P. 
1993 Numerical Recipes in C, (Cambridge: Cambridge)}

\ref{Pumphrey, W.A. \& Scalo, J.M., 1983.  Simulation Models for the Evolution 
of Cloud Systems:  I.  Introduction and Preliminary Simulations.  ApJ, 269, 531.}

\ref{Richtmyer, R.D. 1978, Principles of Advanced Mathematical Physics,
(New York: Springer)} 

\ref{Reid, N., \& Gizis, J. 1997.  Low--Mass Binaries and the Stellar 
Luminosity Function.  AJ, 113, 2246.}

\ref{Reid, N., 1998.  All Things Under the Sun:  The Lower Main--Sequence Mass
Function from an Empiricist's Perspective.  
The Stellar Initial Mass Function, eds. G. Gilmore \& 
D. Howell (San Francisco: ASP).} 

\ref{Roberts, M.S. 1957. The Numbers of Early--Type Stars in the Galaxy and 
Their Relation to Galactic Clusters and Associations. PASP, 69, 59.} 

\ref{Safier, P.N., McKee, C.F., \& Stahler, S.W. 1997.  Star Formation in 
Cold, Spherical, Magnetized Molecular Clouds.  ApJ, 485, 660}

\ref{Sagar, R., \& Griffiths, W.K. 1998. 
Mass Functions of Five Distant Northern Open Star Clusters. 
MNRAS, 299, 777.} 

\ref{Salpeter, E. E. 1955.
The Luminosity Function and Stellar Evolution. 
ApJ, 121, 161.}

\ref{Scalo, J.M. 1986.  The Stellar Initial Mass Function. 
Fundamentals of Cosmic Physics, 11, 1.}

\ref{Scalo, J.M. 1998.  The IMF Revisited:  A Case for Variations. 
The Stellar Initial Mass Function, eds. G. Gilmore \& 
D. Howell (San Francisco: ASP).} 

\ref{Schneider, S. \& Elmegreen, B.G. 1979.  A Catalog of Dark 
Globular Filaments.  ApJS, 41, 87.}

\ref{Shu, F., 1977.  Self--Similar Collapse of Isothermal 
Spheres and Star Formation.  ApJ, 214, 488}

\ref{Shu, F. H. 1985.  Star Formation in Molecular Clouds. 
The Milky Way, IAU Symp. No. 106, eds. H. van Woerden, 
W. B. Burton, \& R. J. Allen (Dordrecht: Reidel).} 

\ref{Shu, F., Adams, F.C., \& Lizano, S. 1987.
Star Formation in Molecular Clouds -- Observations and Theory. 
ARAA, 25, 23. }

\ref{Shu, F., Najita, J., Ostriker, E., Wilkin, F., Ruden, S., \&  
Lizano, S., 1994.  Magnetocentrifugally Driven Flows from 
Young Stars and Disks:  1.  A Generalized Model. ApJ 429, 781.}

\ref{Silk, J. 1995.  A Theory for the Initial Mass Function. 
ApJ, 438, L41.}

\ref{Silk, J. \& Takhashi 1979.  A Statistical Model for the Stellar 
Initial Mass Function. ApJ, 229, 242.}

\ref{Spitzer, L. 1968.  Dynamics of Interstellar Matter and 
the Formation of Stars.  In Nebulae and Interstellar Matter
(Stars and Stellar Systems, Vol. 7), eds. B. M. Middlehurst and L. H. Aller
(Chicago: University of Chicago Press), 1.} 

\ref{Spitzer, L. 1978, Physical Processes in the Interstellar Medium 
(New York:  Wiley--Interscience Press), 282.} 

\ref{Stauffer, J.R., Hartmann, L.W., Prosser, C.F., Randich, S., Balachandran, S., 
Patten, B.M., Simon, T., \& Giampapa, M. 1997.  Rotational Velocities and 
Chromospheric/Coronal Activity of Low--Mass Stars in the Young Open Clusters
IC 2391 and IC 2602. ApJ, 479, 776.}  

\ref{Strom, K.M., Strom, S.E., \& Merrill, M. 1993.  
Infrared Luminosity Functions for the Young Stellar Population
Associated with the L1641 Molecular Cloud. 
AJ, 412, 233.}

\ref{Strom, K.M., \& Strom, S.E. 1994.
A Multiwavelength Study of Star Formation in the 
L1495E Cloud in Taurus. ApJ, 424, 237.} 

\ref{Strom, K.M., Kepner, J., \& Strom, S.E. 1995.
The Evolutionary Status of the Stellar Population 
in the $\rho$ Ophiuchi Cloud Core.  ApJ, 438, 813.} 

\ref{Tohline, J.E., 1980.  The Gravitational Fragmentation 
of Primordial Gas Clouds. ApJ, 239, 417.} 

\ref{Velusamy, T., \& Langer, W. D. 1998. 
Outflow--Infall Interactions as a Mechanism for Terminating 
Accretion in Protostars.  Nature, 392, 685.} 

\ref{von Hippel, T., Gilmore, G., Tanvir, N., Robinson, D., Jones, D.H. 1996.
The Metallicity Dependence of the Stellar Luminosity and Initial Mass Function:
HST Observations of Open and Globular Clusters. .AJ, 112, 192.}

\ref{Walter, F.M., Vrba, R.J., Mathieu, R.D., Brown, A., \& Myers, P.C. 1994.
X--Ray Sources in Regions of Star Formation:  5.  The Low Mass Stars of the 
Upper Scorpius Association.  AJ, 107, 692.} 

\ref{Walter, F.M., Vrba, R.J., Wolk, S.J., Mathieu, R.D., \& Neuhauser, R. 1997.
X--Ray Sources in Regions of Star Formation:  VI.  The R Cr A Association
as Viewed by Einstein.  AJ, 114, 1544.} 

\ref{Wilking, B.A., McCaughrean, M.J., Burton, M.G., Giblin, T., 
Rayner, J., \& Zinnecker, H.  1997. Deep Infrared Imaging of the R Conorae 
Australis Cloud Core.  AJ, 114, 2029.} 

\ref{Williams, D.M., Rieke, G.H., \& Stauffer, J.R. 1995a.
The Stellar Mass Function of Praesepe. ApJ, 445, 359.} 

\ref{Williams, D.M., Comeron, F., Rieke, G.H., \& Rieke, M.J. 1995b.
The Low--Mass IMF in the $\rho$ Ophiuchi Cluster. ApJ, 454, 144.} 

\ref{Williams, D.M., Boyle, D.J., Morgan, W.T., Rieke, G.H., Stauffer, 
J.R., \& Rieke, M.J. 1996.  Very Low Mass Stars and Substellar Objects 
in the Pleiades. ApJ, 464, 238.} 

\ref{Wolfire, M.G., \& Cassenelli, J.P. 1986.  The Temperature Structure in 
Accretion Flows onto Massive Protostars.  ApJ, 310, 207} 

\ref{Zapatero-Osorio, M. R., Rebolo, R., Martin, E. L., \& Garcia Lopez, R.J. 
1996.  Stars Approaching the Substellar Limit in the Alpha Persei Open Cluster. 
A\&A, 305, 519.}

\ref{Zapatero-Osorio, M. R., Rebolo, R., Martin, E. L., Basri, G., 
Maguzzu, A., Hodgkin, S. T., Jameson, R. F., \& Cossburn, M. R., 1997.
New Brown Dwarfs in the Pleiades Cluster. ApJ, 491, L81.}

\ref{Zinnecker, H. 1982.  Prediction of the Protostellar Mass 
Spectrum in the Orion Near--Infrared Cluster. Henry Drapier Sumposium (New York: Ann. New 
York Academy of Science), 395, 226.}

\ref{Zinnecker, H. 1984.  Star Formation from Hierarchical Cloud Fragmentation: 
A Statistical Theory of the Log--Normal Initial Mass Function.  MNRAS, 210, 43.}

\ref{Zinnecker, H., McCaughrean, M. J., \& Wilking, B. A. 1993.  The Initial Stellar Population. 
Protostars \& Planets III, ed. E. Levy \& J. Lunine 
(Tucson: University of Arizona Press), 429.}

\vfill\eject
\null 

\vskip .5in
\centerline{\bf FIGURE CAPTIONS}
\vskip .25in

\caption{Figure 1. \capskip 
Initial mass function for field stars in the solar 
neighborhood taken from a variety of recent studies.
These results have been normalized at 1 $M_{\odot}$.
For both the MS79 and Scalo 86 IMFs we have adopted 
15 Gyr as the age of the Milky Way.  Current work suggests
that the upper end of the IMF ($> 5 M_{\odot}$)
is best represented by a power--law similar to Salpeter (1955)
while the low mass end ($< 1 M_{\odot}$) is flatter 
(Kroupa, Tout, and Gilmore 1993).  The shape 
of the IMF from 1--5 $M_{\odot}$ is highly uncertain. 
See the references listed for details.} 

\caption{Figure 2. \capskip 
Emergent mass distributions for the young clusters
IC 348 (Luhman et al., 1998; Herbig, 1998) and the ONC 
(Hillenbrand, 1997).  Also shown is
the distribution of stellar masses derived 
from the log--normal form of the Miller--Scalo IMF.  Given the 
uncertainties in the PMS tracks we cannot conclude that 
the observed mass distributions are different.  Both 
are broadly consistent with the field star IMF and 
suggest that the IMF below 0.3 M$_{\odot}$ is 
falling in logarithmic units.}

\caption{Figure 3. \capskip 
Ratio of intermediate ($1 M_{\odot} - 10 M_{\odot}$) 
to low ($0.1 M_{\odot} - 1 M_{\odot}$) mass stars for 
star--forming regions as listed in Table I. Also 
shown are the expected distributions if the measurements
were drawn from the Salpeter (1955), Miller \& Scalo (1979), 
or Kroupa, Tout, and Gilmore (1993) mass distributions.
Based on results from the KS test, 
the probability that the observed distributions were drawn 
from these parent populations is $3.28 \times 10^{-8}$, 
0.121, and 0.053 respectively. 
We conclude that the observations 
are inconsistent with the Salpeter IMF extending 
over the range 0.1--10 M$_{\odot}$.}

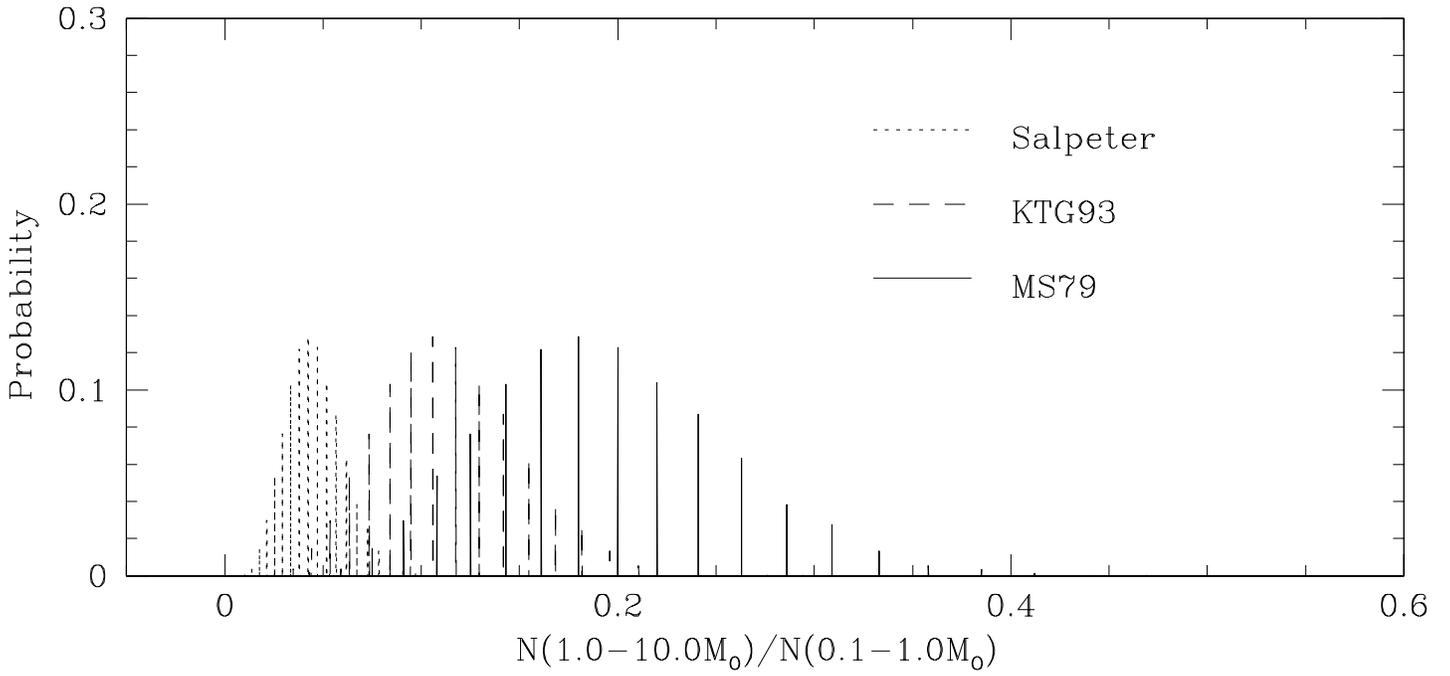
\bye